\documentclass[aps,twocolumn,superscriptaddress]{revtex4}
\usepackage{amsmath,amssymb}
\usepackage{graphics,graphicx}
\usepackage{subfigure}
\usepackage{dcolumn,bm}
\usepackage{psfrag}
\usepackage[dvipsnames]{xcolor}
\usepackage[utf8]{inputenc}
\usepackage{hyperref}
\usepackage{cleveref}
\usepackage{multirow}
\usepackage{braket}
\usepackage{orcidlink}
\usepackage[normalem]{ulem}
\usepackage{soul}
\newcommand{\vc}
{\affiliation{Vidyasagar College, 39, Sankar Ghosh lane, Kolkata 700006, India.}}
\newcommand{\cu}{\affiliation{Department of Physics, University of Calcutta, Kolkata 700009, India.}}
\newcommand{\be}
{\begin{equation}}
	\newcommand{\ee}
	{\end{equation}}

\begin{document}
	\title{Forager with intermittent rest: Better for survival?}
	\author{Md Aquib Molla \orcidlink{0000-0003-0416-1349}}
	\vc
	\author{Sanchari Goswami \orcidlink{0000-0002-4222-5123}}
	\vc
        \author{Parongama Sen \orcidlink{0000-0002-4641-022X}}
        \cu

\begin{abstract}
We study the fate of a forager who searches for food performing a random walk on lattices. The forager consumes the available food on the site it visits and leaves it depleted but can survive up to $S$ steps without food. We introduce the concept of intermittent rest  in the  dynamics which allows the forager to rest with probability $p$ upon consumption of food. The parameter $p$ significantly affects the lifetime of the forager, showing that the intermittent rest can be beneficial for the forager  for 
chosen parameter values. The study of various other quantities reveals interesting scaling behavior with  $p$  and also departure 
from usual diffusive behavior for $0.5 <  p <  1$. In addition to numerical simulations, the problem has been studied with analytical approach in  one dimension and the results up to $p < 0.5$  agree with the numerical ones to a large extent.
\end{abstract}
	
\maketitle

	\section{Introduction}
	Foraging is a habitual tendency for all living organisms, where the forager is in search of a resource. 
	In most cases, a forager moves from one place to another in search of the resource. Food is perhaps the  most common and important resource  without which
    any living organism cannot survive.
    It is also more or less established now  that 
	more than $50000$ years ago, human beings, all initially located 
	in Africa, started 
	spreading to other places, thereby populating the rest of the world  \cite{Stephen}. This migration process is believed to have happened mainly 
	for search of resources. 
	
	In general, for a forager, the fundamental attempt is to maximize the consumption of food. The strategies for the same depend on two main factors: searching the food and then its consumption. The natural tendency of a  forager 
	is  to move to a potentially richer domain as we know from ancient times. This type of foraging problems are  also applicable in several other situations. A few examples include the multiarm bandit problem \cite{Robbins, Gittins}, Feynman’s restaurant problem \cite{Gottlieb}, human memory \cite{Hills, Abbott}, Kolkata Paise Restaurant problem \cite{Chakrabarti}, and so on . Most of these problems, in general, do not consider the depletion of resources. 
	
	Now, in every possible foraging, the basic motivation is not only to find the food, but also to optimize the time for searching. The forager has the urge to survive for long and for that it has to optimize the strategy of foraging.
    The environment through which the forager is moving in search of food, may have heterogeneous or uniform food distribution. In general, the foraging process is studied on a lattice and the food morsels are assumed to be located on the lattice sites. The forager performs a walk on this lattice taking discrete steps and the forager may survive for a certain number of steps after consuming food from one site. Several possibilities for the motion and action of the forager have been considered in the recent past. The simplest case is of course when the forager performs diffusive motion.  In such cases, both the options of taking food
 whenever it encounters a site containing food \cite{Redner2014} or only when its energy falls below a certain threshold value \cite{Redner2018}
 have been considered. Studies have also been made with different approaches in Refs. \cite{Perman, Benjamini, Antal} with consideration of bias towards one side for a one dimensional  walker or forager. In certain cases, the forager may have some sort of intelligence or smartness and its walk is not always a diffusive one when it is close to the food \cite{Redner2022}.  Also, other factors like greed may control the movement \cite{Redner2017}. 
 
There have been a number of experimental and theoretical research works to find out 
 the foraging pattern of animals, especially regarding the issue of whether they perform a L\'evy walk \cite{Edwards,Levy, Dipierro, Klages}. A L\'evy walk mixes many short moves with
 occasional long jumps and thus this type of movement
 can help a forager to reach to the scarce resources more
 efficiently.
 Although some observations do indicate that the forager performs a L\'evy walk, there has been no conclusive general result. 
  The Ornstein-Uhlenbeck foraging process, on the other hand, is less about step lengths and
 more about how an animal moves. Here the movement patterns of foraging animals is somewhat restricted within a defined home range or around a central place  \cite{OUF_Flem1, OUF_Flem2}. It captures the tendency to keep going in the same direction for a while
 which may be called a persistent walk, resulting in piecewise ballistic motion (also called a inertial walk in some references \cite{active_OUF, inertia}). Our model is however not restricted within a home range. It is a diffusive one with another ingredient: intermittent resting. In this model, the forager  relaxes intermittently at a site subject to finding a food morsel at that site.
The forager here can be regarded as making steps of different length 
over the same interval of time somewhat similar to a L\'evy walk.
	
  In this paper, we consider the possibility of
 intermittent resting which may be beneficial for the forager in two senses:
The food gets conserved for longer time and the forager lives longer. 
 We have introduced a new parameter in the dynamics and explored its effect on 
 various relevant quantities. 
 Section \ref{desc} provides a detailed description of the model. The results and conclusions are presented in the next two sections. 

	\section{Model Description}
	\label{desc}

    In the model considered in the present work, the forager is a simple random walker that moves through the environment in search of food and depletes the resource upon consumption. The environment is a $d$-dimensional lattice (typically $d$=$1$ or $2$), where food is initially distributed uniformly, each lattice site contains exactly one morsel of food. If the walker lands on a site, then containing food, it consumes the food completely and becomes ``full,” which allows it to continue moving for a predefined number of additional steps, denoted by $S$, without needing to eat again \cite{Redner2014}. The walker's metabolic capacity, or starvation time, is therefore $S$; that is, the walker may take up to $S$ consecutive steps surviving without encountering food, after which it  dies if food is not available in the next step. 
    If the walker visits an empty site, then its internal energy decreases by one unit.
    
	
	The key feature considered in our model is  intermittent rest which is  incorporated through  a parameter $p$. Here $p$ denotes  the probability that the forager  takes rest at a site where it  gets food and consumes it.
 
The forager starts from a point which we denote as the  origin of a $d$-dimensional lattice. 
    As all the sites contain a  food morsel in the beginning, it will find and consume the food at the origin.  However, as there is an intermittent resting probability, the forager can either take rest (with probability $p$) or jump off (with probability $1-p$) to an adjacent site. 
	As  the forager can survive for $S$ steps/time units  after consuming 
food (without consuming any further food morsel), for the extreme limit $p=1$,  the forager stays forever at the origin which  will obviously lead to its  death
	and   the life-time of the forager in this case is $\tau = S$. 
    However, for $p < 1$, the  forager can  
       move  out of that site and continues to  move as a simple random walker until landing on a site where food is available.


    Consider a forager  in a one-dimensional lattice with $S=2$ and $p=0.5$.
    \begin{itemize}
        \item At $t=0$, it starts from $x=0$, consuming the food there.
        \item At $t=1$ and $t=2$, it can move with probability $0.5$, and suppose it moves to $x=1$ and $x=2$ respectively, consuming the food at each site.
        \item At $t=3$, suppose it moves to $x=1$, where there is no longer any food. So it is bound to move again.
        \item At $t=4$, it may either go to the origin or $x=2$, both sites are empty. It will die as $S=2$ and it has spent two steps without food and therefore lifetime $\tau=4$.
    \end{itemize}

    In case the forager takes rest at time $t = 2$ after reaching $x = 1$ at $t = 1$ and then move in a similar fashion as described above, its lifetime will be $\tau = 5$.

	This probabilistic intermittent resting introduces a significant element of randomness in the forager's path and affects both its movement strategy and its survival time. The forager's lifetime depends on its ability to find food as it moves across the lattice, as well as on the impact of resting probability. However, resting too often could lead to starvation if it fails to find new food sources in time.
	
	Our model helps to understand how the balance between movement and intermittent rest affects the forager's survival and the distribution of its lifetime. By varying $p$ and $S$, we explore the conditions under which the forager survives for a longer or shorter period. The average lifetime $\tau$, the total number of distinct sites, 
    averaged over all realisations, visited by the forager $N$ and the relaxation time $T(x)$, i.e., the resting time of the forager at site $x$ are studied. In addition, we have calculated the distribution of $P(N)$ of $N$, the inter-encounter time $N_t$, and its corresponding distribution $P_{t}$ which will be discussed in the following sections.

	\section{Results}
	
	In this section,  the results for the forager with intermittent resting are presented, where we consider the
  dynamics  up to the point when the forager died due to starvation. Initially, in a $d$-dimensional ($d=1$ or $2$) lattice grid, all sites contain food. The forager starts from the origin.
    As mentioned earlier, it may so happen that the forager stays at its initial position (origin) and never comes out of it. Finally, it starves to death and correspondingly the lifetime $\tau = S$. Of course, this would be the lower bound of the lifetime. Otherwise, it will continue moving through the lattice. Whenever it steps on a site containing food, the food will be consumed and the corresponding site will become empty. In this way, after a certain number of time steps, there will be a considerable number of empty sites. This cluster of empty sites can be called a desert which is depleted of food. As the forager gradually eats the food morsels, the size of desert is enlarged. When the size of the desert becomes larger than a threshold limit $L_c$ (detailed analysis is in Ref. \cite{Redner2016}) the forager is unable to cross the desert. As a consequence, the forager starves to death. 
	
	
	We approach the problem using both analytical and numerical methods for the $1d$ case. The detailed analysis using the analytical method  is presented
in   the appendix where we solve the diffusion equation for the forager. 
    For a particular configuration, at some time step, the forager is dead and we calculate the number of distinct sites $\eta$ visited by it before its death. Finally, we calculate the average number of distinct sites $N$
    given by
  \begin{equation}
		N = \sum_{\eta} \eta \Pi(\eta),
    \end{equation} 
  where $\Pi(\eta)$ is the probability distribution of $\eta$.

    The probability that a random walker has visited $N$ average number of distinct sites in $1d$ is given as:
	\begin{equation} \label{P_N}
		P(N) = Q_2 Q_3 Q_4 \cdots Q_{N}(1-Q_{N+1}),
	\end{equation} 
	where $Q_j = \int_0^S F_j(t) dt$ and $F_j(t)$ is the probability of the walker reaches either end of an interval of length $ja$ (where $a$ is the lattice spacing) at step $t$ when starting from a distance $a$ from one end. Each $Q_j$ represents the probability contribution from the interval expanding in length from $j-1$ to $j$, as the walker successfully reaches one of the boundaries within $S$ steps. Meanwhile, the term $Q_{N+1}$ corresponds to the final excursion, during which the walker fails to reach a boundary and consequently starves. Scaling $N$ by $\theta$ where $\theta = aN/(\pi\sqrt{DS})$, we get, $P(N)$ as follows (details mentioned in Appendix A): 



\begin{widetext}
\begin{multline}
P(N) = \frac{4}{\pi} \sum_{m = 1}^{\infty} \frac{\sin{\left( \frac{(2m+1)\pi}{N+1}\right)}}{(2m+1)}
	\times\frac{\gamma\left[ \left(\frac{(2m+1)\pi }{(N+1)a}\right)^2D\Theta p\zeta + 1 , \left(\frac{(2m+1)\pi }{(N+1)a}\right)^2DS\right]}{\left[\left(\frac{(2m+1)\pi }{(N+1)a}\right)^2D\right]^{\left(\frac{(2m+1)\pi }{(N+1)a}\right)^2D\Theta p\zeta}} \\
	\times \exp{\left[ \sum_{2 \leqslant j \leqslant N} \ln{\left[ 1 - \frac{4}{\pi} \sum_{n = 1}^{\infty} \frac{\sin{\left( \frac{(2n+1)\pi}{j}\right)}}{(2n+1)}\frac{\gamma\left[ \left(\frac{(2n+1)\pi }{ja}\right)^2D\Theta p\zeta + 1 , \left(\frac{(2n+1)\pi }{ja}\right)^2DS\right]}{\left[\left(\frac{(2n+1)\pi }{ja}\right)^2D\right]^{\left(\frac{(2n+1)\pi }{ja}\right)^2D\Theta p\zeta}}  \right]}\right]}, \label{generalPN}
\end{multline}
\end{widetext}
where $D$ is the diffusivity and $\gamma (s,x)= \int_x^{\infty} t^{s-1}e^{-t} dt$ is the incomplete $\gamma$ function. 
	The average lifetime $\tau$ is given by, 
	\begin{equation}
		\tau = \sum_{j \geqslant 1} \left( \tau_1 + \tau_2 + \cdots + \tau_j + S \right)P(j).
	\end{equation}
    Here $\tau_j$ denotes the mean time taken by the walker to reach either boundary of the interval during the $j$th excursion, given that the walker reaches a boundary before starving. The factor $S$ represents the contribution from the final excursion. By definition
	  \begin{equation} \label{tau_j}
	      \tau_j = \frac{\int_0^S t F_j(t) dt}{\int_0^S F_j(t) dt}.
	  \end{equation}
      It is to be noted that, here we  write Eq. \ref{tau_j} regarding time as a continuous variable in the large $S$ limit, when the discrete intervals are very small compared to $S$. The function $F_j(t)$ appearing in Eq. \ref{tau_j} is shown explicitly in the appendix [see Eq. \ref{eqn: F_j}].

 The average number of distinct sites $N$, its distribution $P(N)$, the average lifetime $\tau$ and other relevant quantities have been studied numerically and will be discussed in detail in the following subsections.
 The analytical result for $P(N)$ and other quantities in $1d$ are also discussed.

	\subsection{Average lifetime $\tau$}\label{A}

\begin{figure}[h!]
		\begin{center}
			\includegraphics[angle=-90, trim = 0 150 0 150, clip = true, width=0.99\linewidth]{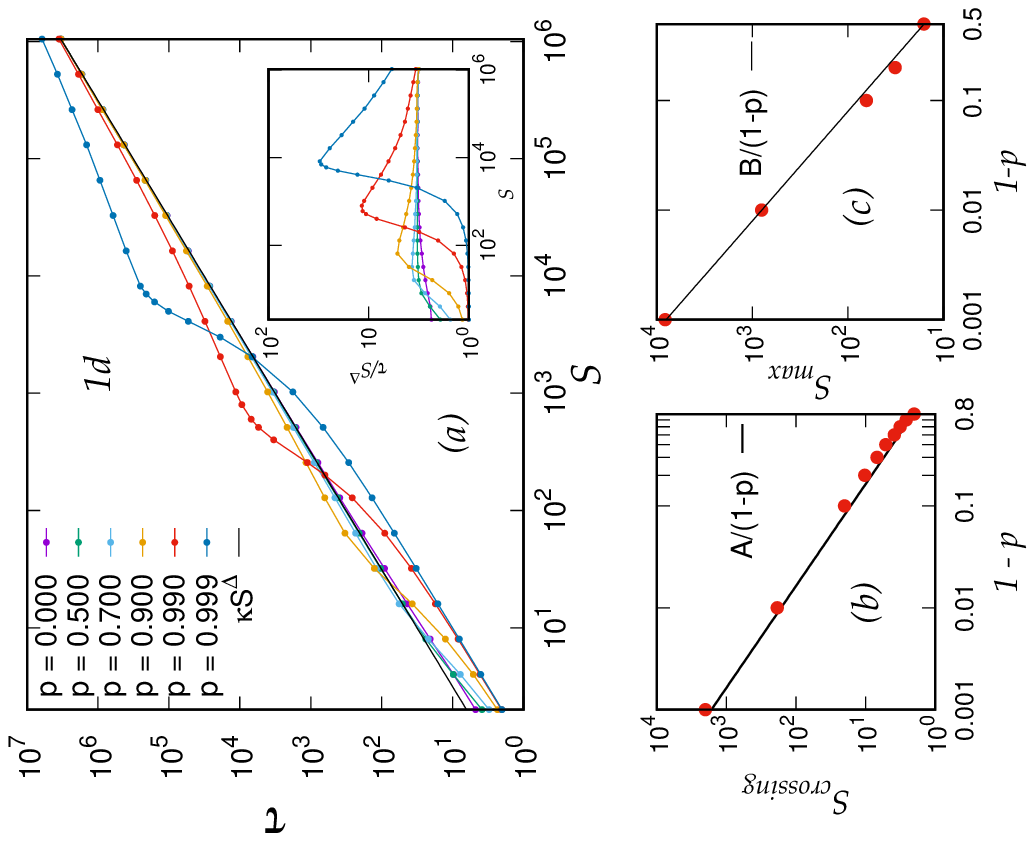}
			\caption{(a) Lifetime $\tau$ against starvation time $S$ for $0 \leqslant p < 1$ in 1$d$. Power law behavior is observed  for $S \gg S_{\text{crossing}}$ as $\tau \sim \kappa S^{\Delta}$, with $\Delta \simeq 1.0$ and $\kappa \simeq 3.18$. In the inset, the behavior for the scaled lifetime $\tau/S^{\Delta}$ is shown. 
            The deviation from the conventional scaling is observed over a larger range of values of $S$ as $p$ increases and its  maximum value occurs at $S = S_{\text{max}}$, with $S_{\text{max}}$ increasing with $p$. (b) $S_{\text{crossing}}$ is plotted against $1-p$ and the curve is dictated by $S_{\text{crossing}} = \frac{A}{1-p}$, where $A = 1.64 \pm 0.03$. (c) $S_{\text{max}}$ as a function of $1-p$ is shown; $S_{\text{max}} = \frac{B}{1-p}$ with $B=7.98 \pm 0.12$.}
		\label{1D_Tave}
	\end{center}
\end{figure}

The total lifetime $\tau$ of a forager, including its times of rest on the same site, is calculated, 
before it starves to death inside the desert created by consuming the food morsels.  We run the simulation over $10^6$ realizations and calculate the average lifetime $\tau$. The average lifetime is studied against $S$ for different $p$ values ($0 \leqslant p \leqslant 1$). The results for $\tau$ for the 1$d$ case  has been shown in Fig. \ref{1D_Tave}(a).
From Fig. \ref{1D_Tave}(a), it is clear that for any $0 < p < 1$, the curves start below the $p=0$ curve, cross the $p=0$ curve at some $S$ value, denoted as $S_{\text{crossing}}$, and then go well above the $p=0$ curve and reach a maximum. In the inset of Fig. \ref{1D_Tave}(a), the scaled lifetime $\tau/S^{\Delta}$ with $\Delta \simeq 1.0$ is shown as a function of $S$, where these maxima for different $p$ can be clearly seen. This ensures an increase in lifetime. For very high $S$, the $\tau$ versus $S$ curves asymptotically approach the $p=0$ curve.
The following points are also clear from Fig. \ref{1D_Tave}(a):
\begin{itemize} 
	\item For low $p$, the curves start (low $S$) just below the $p=0$ curve. As $p$ is made higher, the starting point of the curves becomes lower.
	\item $S_{\text{crossing}}$ is less (larger) for lower (higher) values of $p$. The variation of $S_{\text{crossing}}$ is shown in Fig. \ref{1D_Tave}(b).
       \item For $S \gg S_{\text{crossing}}$ the behavior for 1$d$ can be described as: 
	\begin{equation} \label{tau}
		\tau \propto S^{\Delta},
	\end{equation}
	with $\Delta \sim 1.0$.
    \item The value of $S$ at which $\tau/S^{\Delta}$ shows maximum for a particular $p$, denoted as $S_{\text{max}}$, shows similar scaling behavior as that of $S_{\text{crossing}}$ and is shown in Fig. \ref{1D_Tave}(c).
\end{itemize}

For  $p=0$, the lifetime $\tau \propto S$ in 1$d$, a known result  \cite{Redner2014}
which is obtained correctly from the present simulations shown in Fig. \ref{1D_Tave}(a). For $p = 1$, since the forager stays at the origin until starvation (no movement), it indicates $\tau = S$ as well. For the intermediate values of $p$, a transition from $p = 1$ line to $p = 0$ line can be observed in  Fig. \ref{1D_Tave}(a). In these cases, due to intermittent resting, the forager partially retains its position and consumes less amount of food compared to the $p=0$ case. For low $S$ the forager can survive for less time with one food morsel. As its position is restricted  due to intermittent resting, its lifetime becomes less. For very high $S$, however, it can live  long  once it gets food, and therefore even if its motion is somewhat restricted, it can survive longer. 
For higher $p$ and higher $S$, the increase in lifetime can be explained  in the following way: 
the forager with large $S$ can go without food for a long period. Even with a high $p$ value, due to which it tends to stay for a longer time at any location, it has chances to live. Thus, as the forager does not move much, the sites where the forager does not reach by that time, can remain filled. This happens with a higher probability for larger and larger $p$ as effectively the forager stays at a site for a longer and longer time. Thus a  smaller desert is created for high $p$ and $S$. We will come back to this issue of  extension of lifetime after describing other results in Secs. \ref{B} and \ref{C}.
For very high $S$, as the behavior merges with that of the $p=0$ case, $\tau \sim S^{\Delta}$ (with $\Delta \simeq 1$) is encountered, independent of the value of $p$. 

\begin{figure}[h!]
	\begin{center}
		\includegraphics[angle=-90, trim = 0 150 0 150, clip = true, width=0.99\linewidth]{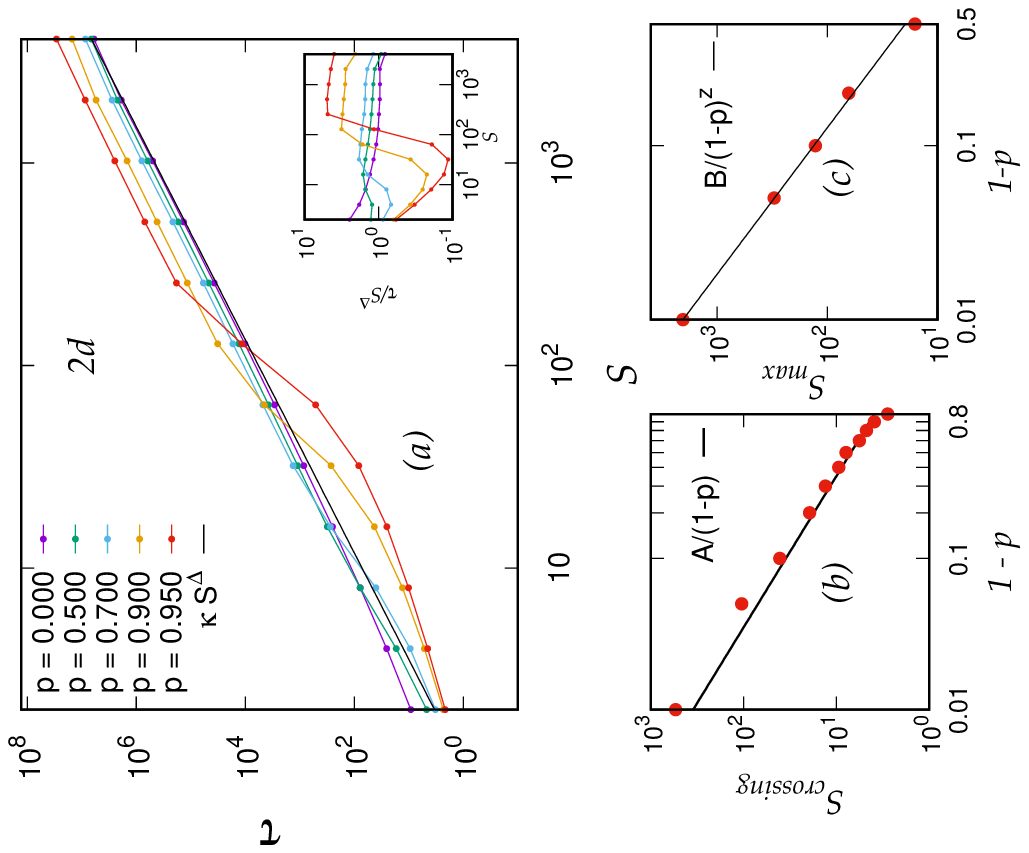}
  
        \caption{(a) Lifetime $\tau$ against starvation time $S$ with $0 \leqslant p < 1$ in $2d$. Similar power law behavior as in Fig. 1 (a) is observed with the exponent $\Delta \simeq 1.9$ and $\kappa \simeq 0.91$. (b) $S_{\text{crossing}}$  against $1 - p$. It has been observed that $S_{\text{crossing}} = \frac{A}{1 - p}$ where $A = 3.52 \pm 0.19$.(c) $S_{\text{max}}$ as a function of $1 - p$. $S_{\text{max}} = \frac{B}{(1 - p)^z}$ with $B = 8.60 \pm 0.32$ and $z \simeq 1.19$.}
        \label{2D_Tave}
	\end{center}
\end{figure}


In the $2d$ case, the forager experiences greater freedom of movement, allowing it to explore more directions before encountering previously visited or depleted sites. This increased mobility leads to a higher value of the scaling exponent, with $\Delta \simeq 1.9$ as shown in Fig. \ref{2D_Tave}(a) for high $S$ for any $p$. For $p=0$, the value of $\Delta$ in two dimensions is mentioned as 2.0 in Ref. \cite{Redner2014}.
 The overall behavior of the curves for different values of $p$ closely resembles the patterns observed in the one-dimensional case, indicating that the foraging dynamics remain qualitatively similar across dimensions.  It is to be mentioned that just like the $1d$ case, here also the extension of lifetime is observed for high $p$ and $S$.

The behavior of $S_{\text{crossing}}$ is shown in Fig. \ref{2D_Tave}(b). In the inset of Fig. \ref{2D_Tave}(a), $\tau/S^{\Delta}$ versus $S$ for different $p$ is shown. The corresponding scaling of $S_{\text{max}}$ is shown in Fig. \ref{2D_Tave}(c).

The variation of $S_{\text{crossing}}$ for both $1d$ and $2d$ are found to be of the form : 
\begin{equation} \label{Scrossing}
	S_{\text{crossing}} = \frac{A}{1-p},
\end{equation}    
and that of $S_{\text{max}}$ is close to
\begin{equation}
    S_{\text{max}} = \frac{B}{(1-p)^z},
\end{equation}
where $z=1.0$ in 1$d$ and $z \simeq 1.19$ in 2$d$.
These variations are shown in Fig. \ref{1D_Tave}(b) and \ref{1D_Tave}(c) for 1$d$ and in Fig. \ref{2D_Tave}(b) and \ref{2D_Tave}(c) for 2$d$.

It is to be noted that although Eq. \ref{tau} is given for the asymptotic limit, for very low $S$, $\tau$ shows the same behavior, i.e., $\tau \propto S^{\Delta}$ for both 1$d$ and 2$d$.

\subsection{Average number of distinct sites $N$}\label{B}
\begin{figure}[h!]
	\begin{center}
		\includegraphics[angle=-90, trim = 0 0 0 0, clip = true, width=0.99\linewidth]{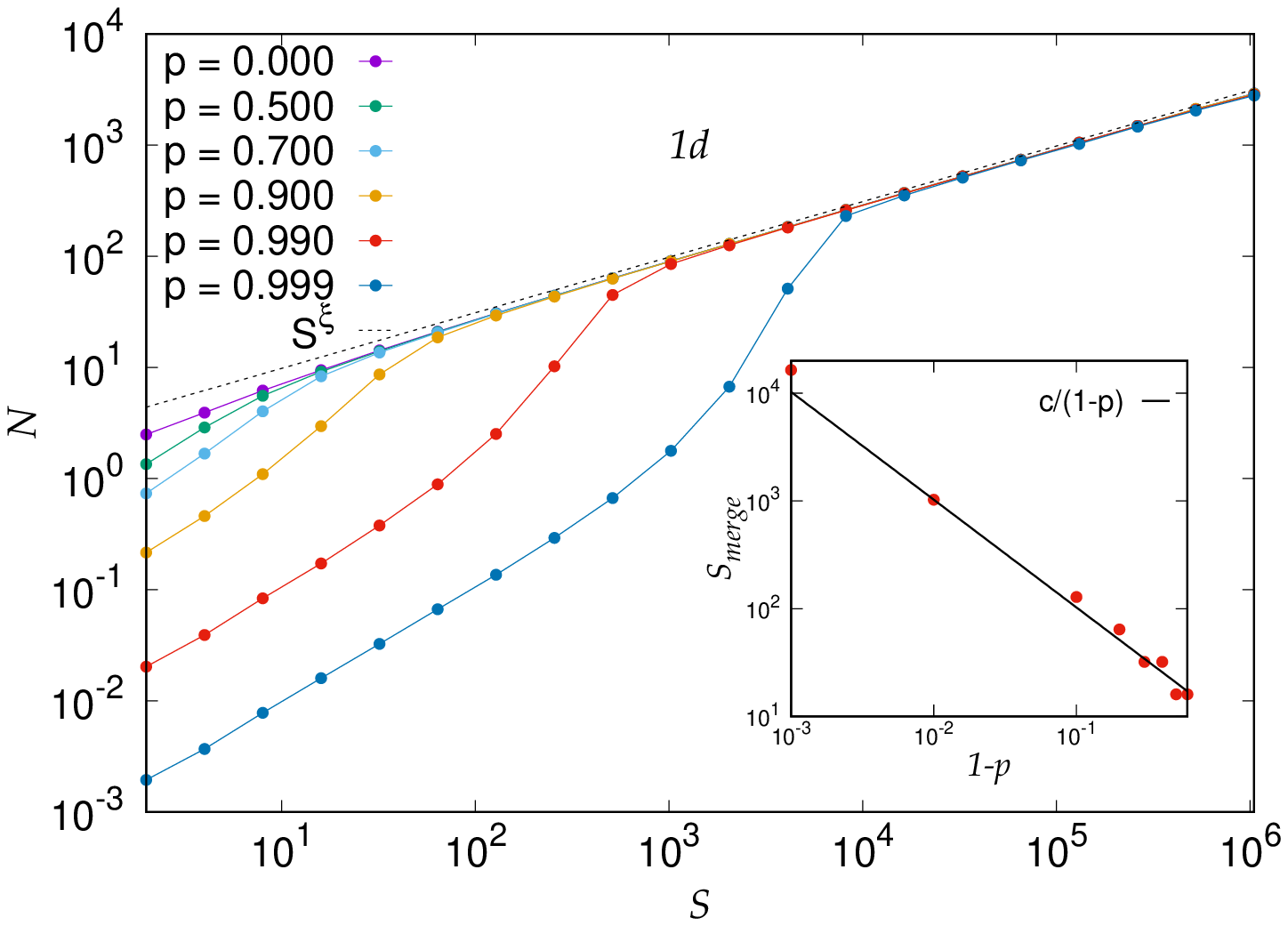}
		\caption{Behavior of average number of distinct site $N$ as a function of starvation time $S$ with $0 \leqslant p < 1$. For $S \gg S_{\text{merge}}$, power law behavior is observed, the exponent being $\xi \simeq 0.50$. 
        $S_{\text{merge}}$ against $p$ has been shown in the inset, which follows a relation $S_{\text{merge}} =  \frac{c}{1-p}$ where, $c =10.26 \pm 0.11$.} 
		\label{1D_N}
	\end{center}
\end{figure}

\begin{figure}[h!]
	\begin{center}
		\includegraphics[angle=-90, trim = 0 0 0 0, clip = true, width=0.99\linewidth]{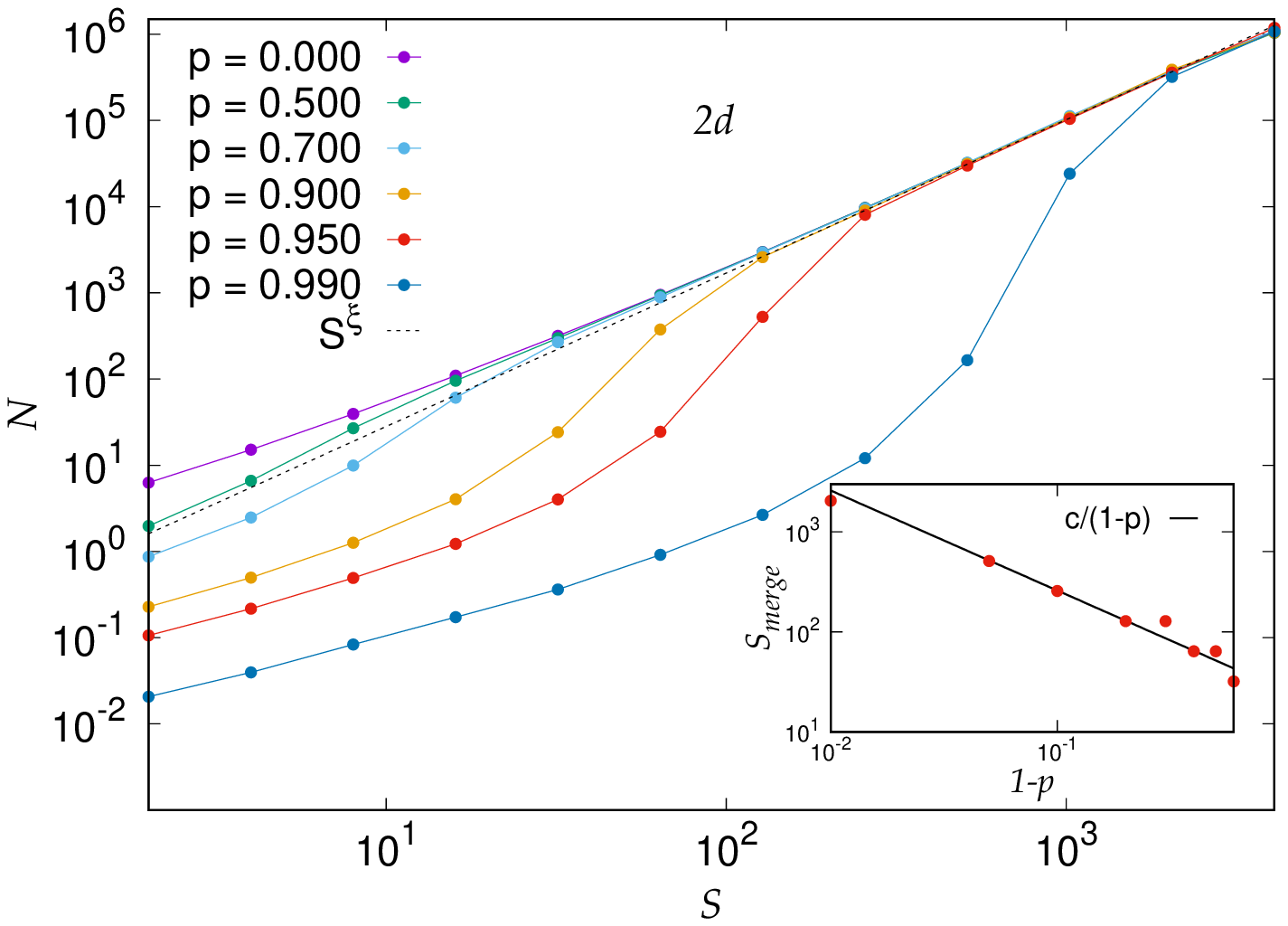}
		\caption{Average number of distinct site $N$ against starvation time $S$ in $2d$ with $0 \leqslant p < 1$. For $S \gg S_{\text{merge}}$, $\xi \simeq 1.78$. $S_{\text{merge}} = \frac{c}{1-p}$ is shown in the inset with $c = 25.91 \pm 0.84$.} 
 		\label{2D_N}
	\end{center}
\end{figure}

The average number of distinct sites, $N$ against $S$ with $p$ as a parameter in $1d$ and $2d$ are studied and the results are shown in Fig. \ref{1D_N} and Fig. \ref{2D_N} respectively.

In the asymptotic limit, i.e. for large $S$, for any $p \neq 0$, it has been observed that, the average number of distinct sites $N$ shows power law behavior with $S$ and approaches the $p=0$ curve. 

It is known that, for the unrestricted walker, the average number of distinct sites visited $N \simeq \sqrt{8S/\pi}$ \cite{Weiss}. 
In Ref. \cite{Redner2014}, it was shown that $N$ goes as $\sqrt{S}$. Our results for $p=0$, both numerical and analytical, matches with those in Ref. \cite{Redner2014}. In general, for all $p$, 
\begin{equation} \label{N}
	N \propto S^{\xi}
\end{equation}
where $\xi \simeq 0.50$ in $1d$ for the asymptotic limit. 
However, in the small $S$ region, as $p \rightarrow 1$, the curves show linear behavior in the log-log plot indicating $\xi \simeq 1$ there. In the intermediate region, it shows some non-linearity and finally merges with $S^{\xi}$ at $S = S_{\text{merge}}$. $S_{\text{merge}}$ thus can be defined as the particular $S$ for a chosen $p \neq 0$ beyond which the curves for average number of distinct sites visited $N$ coincide with the $p=0$ curve. In the inset of Fig. \ref{1D_N}, we have shown the variation of $S_{\text{merge}}$ with $1-p$, and found that $S_{\text{merge}}$ follows the relation,
\begin{equation} \label{Smarge}
	S_{\text{merge}} = \frac{c}{1-p}
\end{equation}
where $c =10.26\pm 0.11$ shown  in Fig. \ref{1D_N}. The increase of $S_{\text{merge}}$ with $p$ is also due to larger restriction for higher $p$ values.  

In the $2d$ case, the forager has more freedom to access the food morsels, causing increment of average number of distinct sites. From our numerical result, we obtain $\xi \simeq 1.78$. For $S_{\text{merge}}$, $c = 25.91 \pm 0.84$. For $p \rightarrow 1$, for low $S$ and very high $S$, $N$ has the behavior $S^{\xi}$ where $\xi$ is approximately $1.78$.

It is to be noted that the $p = 1$ case is trivial because the forager stays at the origin until it starves to death i.e $N = 1$, and therefore we have not shown it in Fig. \ref{1D_N} and Fig. \ref{2D_N}. 

For both $1d$ and $2d$, we can observe from Fig. \ref{1D_N} and Fig. \ref{2D_N} one feature in common: For high $S$, for any $p$, the average number of distinct sites visited $N$ is never crossing the $p=0$ curve; it  ultimately merges with the $p=0$ curve at $S = S_{\text{merge}}$ from a lower value. Inclusion of intermittent resting makes the number of visited sites smaller. The average number of distinct sites visited $N$ with resting is a strict subset of those that would have been visited without resting. Thus inclusion of resting means the $p=0$ curve is approached from below.

As already mentioned in Sec. \ref{A},  for high $p$ and $S$, the desert is not large enough and there are plenty of sites containing food which helps the survival of the forager. The indication is clear from Fig. \ref{1D_N} and \ref{2D_N} also.

\begin{figure}[h!]
	\begin{center}
		\includegraphics[angle=-90, trim = 0 0 0 0, clip = true, width=0.99\linewidth]{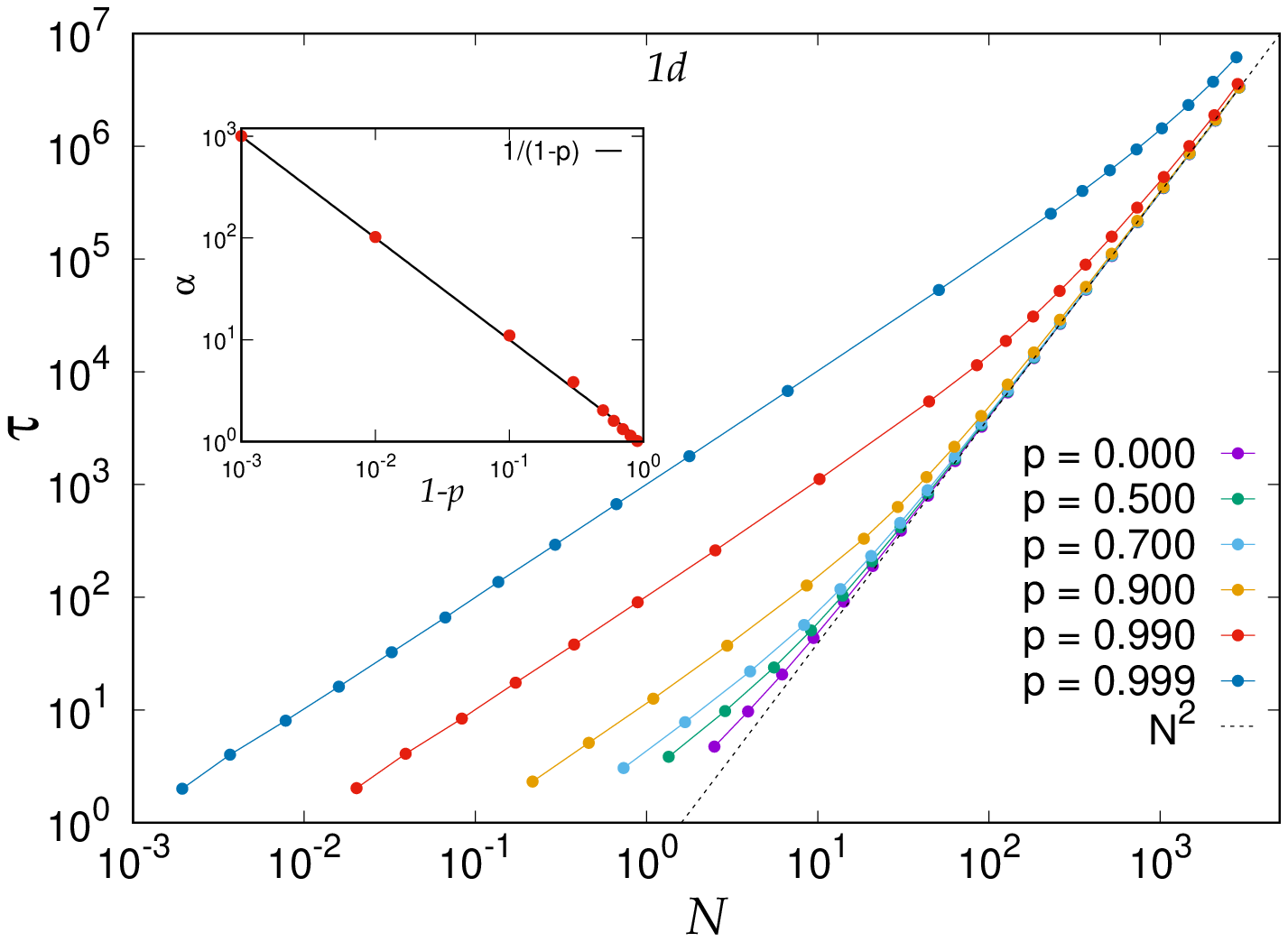}
		\caption{Variation of lifetime $\tau$ with average number of distinct site $N$ for $1d$. The variation follows the relation $\tau(N) = \alpha(p) N$ in small $N$ region for $p<1$. In the large $N$ region, the curve goes as $N^2$. In the inset, the variation of $\alpha$ with $1-p$  is shown as  $\alpha = \frac{1}{1-p}$.}
		\label{1D_T_N}
	\end{center}
\end{figure}

\subsection{Relation between $\tau$ and $N$}\label{C}
\begin{figure}[h!]
	\begin{center}
		\includegraphics[angle=-90, trim = 0 0 0 0, clip = true, width=0.90\linewidth]{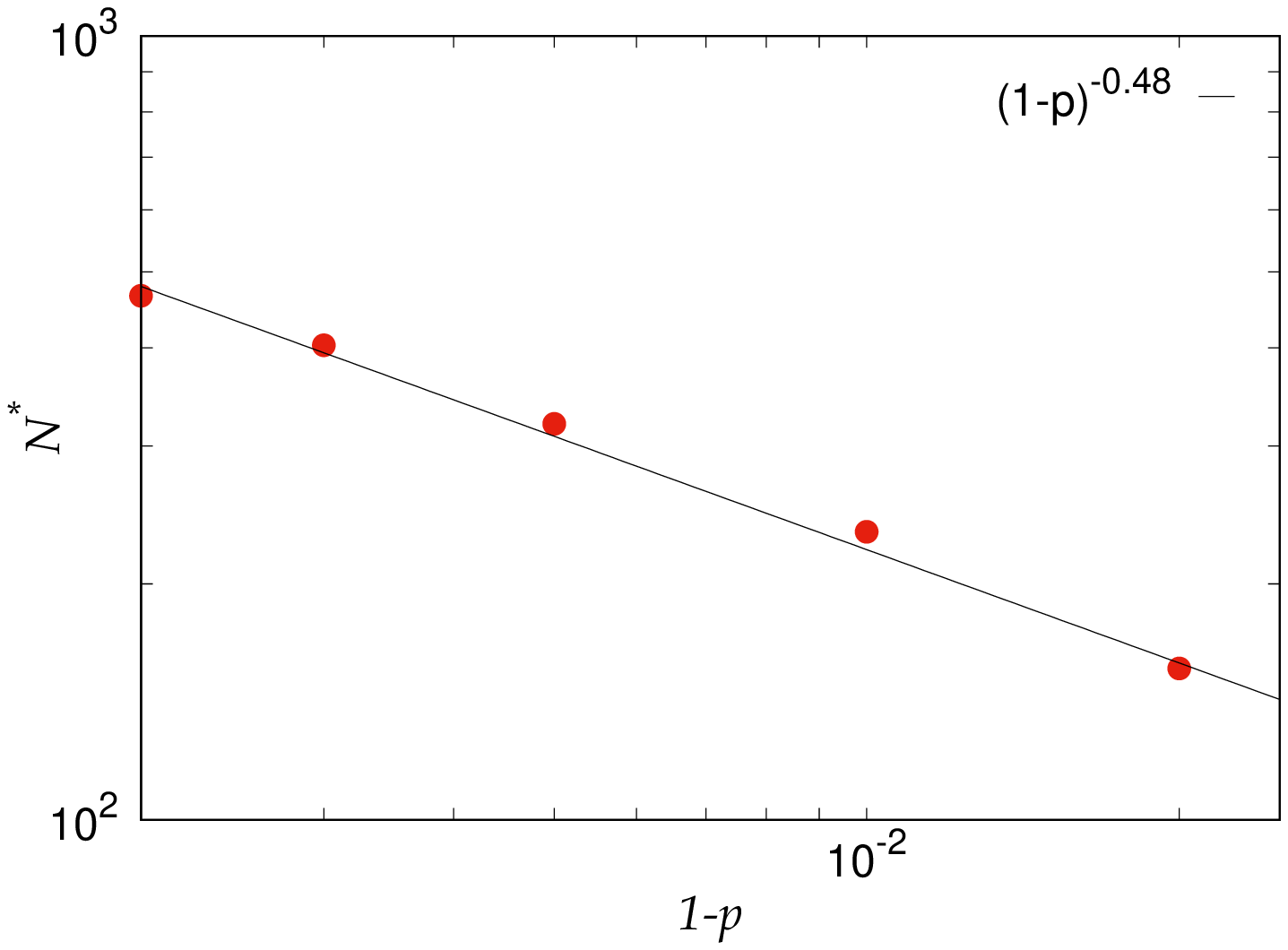}
		\caption{The scaling behavior of $N^*$ against $1-p$. It is observed that $N^* \propto \frac{1}{(1-p)^{0.48}}$.}
		\label{Nstar_vs_p}
	\end{center}
\end{figure}

Now, we would like to see the relation between $\tau$ and $N$. We used a heuristic approach here. Combining Eq. \ref{tau} and Eq. \ref{N} we get the following : 
\begin{equation} \label{tau_N_co}
 \tau \propto N^{\Delta/\xi}.   
\end{equation}
 This is in agreement with the results presented in Fig. \ref{1D_T_N} for 1$d$. Here, for any $p < 1$, the $\tau$ versus $N$ curves show a crossover behavior from $N$ to $N^2$ for 1$d$ as $N$ is increased. For $p>0$ in 1$d$, from the high $N$ region ($S$ is also high there), we have $\tau \sim N^2$ as $\Delta=1.0$ and $\xi=0.5$ there (already discussed in Sec. \ref{A}). For low values of $N$, however, the behavior of $\tau$ is observed as $\tau \sim N$. This is also clear from the Fig. \ref{1D_Tave} and \ref{1D_N} and from the corresponding exponents. The particular $N$ beyond which a crossover behavior of $\tau$ from $N$ to $N^2$ is observed, is termed as $N^*$. This is shown in Table \ref{Nstar}. In the same table, the third column shows the $N$ values corresponding to $S_{\text{merge}}$ as found from Fig. \ref{1D_N}. The variation of $N^*$ with $p$ is shown in Fig. \ref{Nstar_vs_p}.
 
 It is clear from Fig. \ref{1D_T_N} that for high $p$ and low $N$ the increase in lifetime is ensured. This means by using limited amount of resource the forager can survive for longer time, as intermittent rest is included. 
 
The lower $N$ region of the curves are fitted with,
\begin{equation} \label{tau_N}
	\tau(N) = \alpha(p) N
\end{equation}

\begin{table}[h!]
\begin{center}
    \begin{tabular}{|c|c|c|}
        \hline
        $p$ & $N^*$ & $N$ at $S_{\text{merge}}$ \\
        \hline 
        \hline
        $0.500$ & $8$ & $9$\\
        $0.700$ & $12$ & $13$\\
        $0.800$ & $16$ & $20$\\
        $0.900$ & $31$ & $29$\\
        $0.990$ & $233$ & $85$\\
        $0.999$ & $1195$ & $230$\\
        \hline
    \end{tabular}
\end{center}

\caption{$N^*$ values for different $p$ is shown. In the third column the value of $N$ at $S_{\text{merge}}$ is shown for comparison.}
\label{Nstar}
\end{table}

The variation of $\alpha(p)$ is shown in the inset of Fig. \ref{1D_T_N}, where $\alpha$ follows the relation $\alpha(p) \approx \frac{1}{1-p}$. 

Although for 2$d$, there is no crossover behavior and for all $N$, we have approximately $\tau \sim N$, as expected from the heuristic approach, for $p>0$, the lifetime is still increased.

\subsection{Distribution of average number of distinct site $P(N)$}
\begin{figure}[h!]
	\begin{center}
		\includegraphics[angle=-90, trim = 0 0 0 0, clip = true, width=0.99\linewidth]{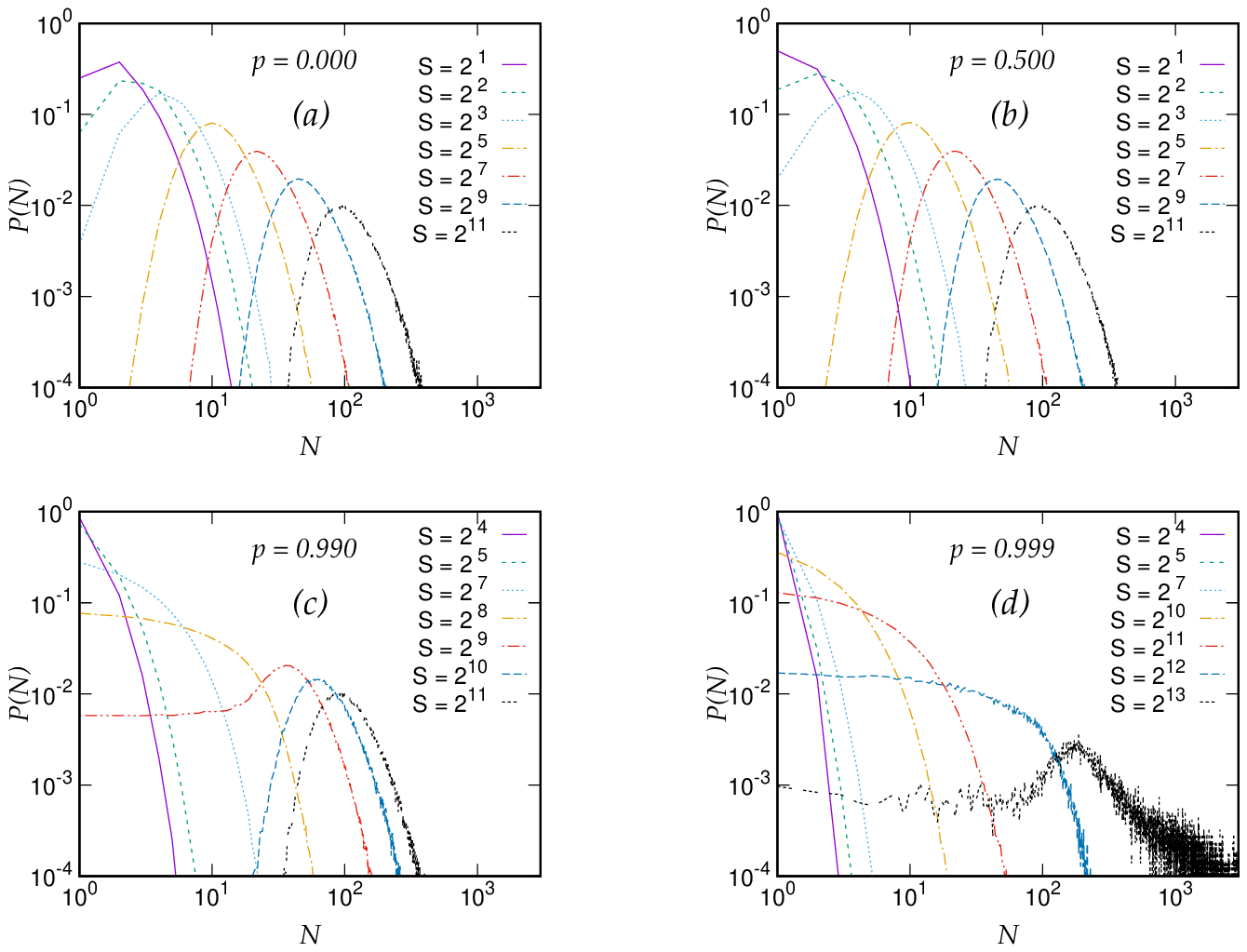}
		\caption{Distribution $P(N)$ of average number of distinct site $N$ (in $1d$ case) for (a) $p = 0.000$, (b) $p = 0.500$, (c) $p = 0.990$ and (d) $p = 0.999$.}
		\label{1D_PN}
	\end{center}
\end{figure}

We study the distribution of average number of distinct sites $N$, denoted as $P(N)$. In $1d$ case, we have run $10^6$ realizations for $p = 0.000, 0.500, 0.990$ and $0.999$ and $S = 2^1$ to $2^{13}$. This is shown in Fig. \ref{1D_PN}. For $p=0$ (Fig. \ref{1D_PN}a) (a), the curves show a dome like structure where the peak position moves with $\sqrt{S}$ in $1d$ which agrees with the results presented in Fig. \ref{1D_N}. The similar pattern also appears in $p = 0.500$ for case (b).
Now, as we increase $p$, $S_{\text{merge}}$ plays some interesting role. For $p = 0.990$ and $S < S_{\text{merge}}$ the dome shape has been lost [Fig. \ref{1D_PN}(c)]. For $S \geqslant S_{\text{merge}}$ the dome shape reappears. For $p = 0.999$, the dome is only visible for $S \geqslant S_{\text{merge}}$ where $S_{\text{merge}}$ is as large as $2^{13}$ Fig. \ref{1D_PN}(d).

\begin{figure}[h!]
	\begin{center}
		\includegraphics[angle=-90, trim = 0 0 0 0, clip = true, width=0.99\linewidth]{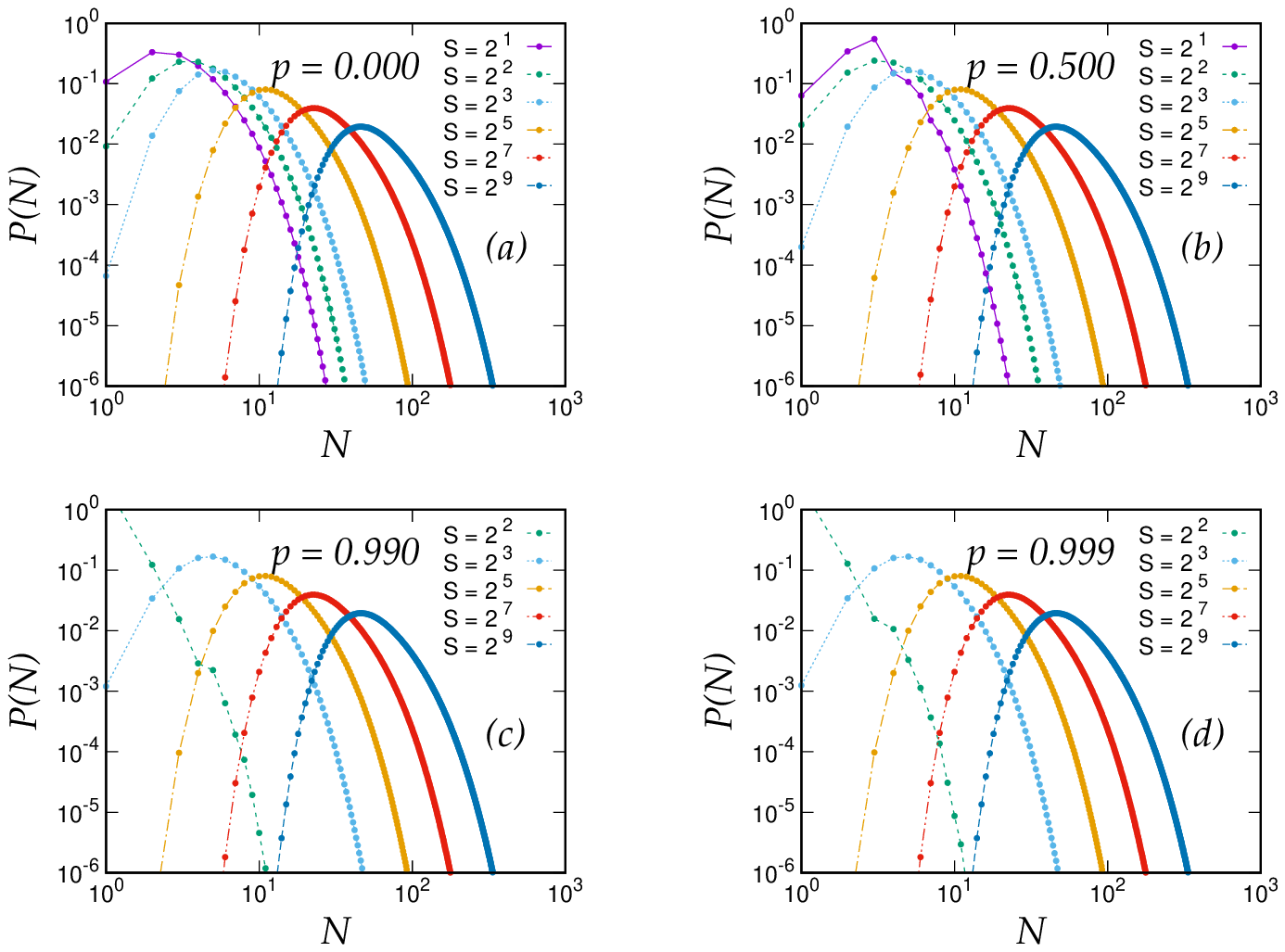}
		\caption{In $1d$, the distribution $P(N)$ of average number of distinct site $N$ from analytical approach for (a) $p = 0.000$, (b) $p = 0.500$, (c) $p = 0.990$ and (d) $p = 0.999$. For $p> 0.5$. the results show a significant difference from those obtained for numerical simulations.}
		\label{Analytical_1D_PN}
	\end{center}
\end{figure}

The distribution has been studied analytically starting from a diffusion equation which is shown in Appendix A in Eq. \ref{DiffusionEquation}. The corresponding distribution function is shown in Eq. \ref{generalPN}. The analytical result matches exactly with the result of the paper \cite{Redner2014} by putting $p=0$. However, for $p \neq 0$, the analytical result and numerical results show agreement for low values of $p$, up to  $p=0.5$. This may be due to the fact that the analytical result contains several factors that have to be approximated (see Appendix A for details). In particular, we have taken $p$-dependent terms in the diffusion equation [Eq. \ref{DiffusionEquationA2}] which involves several approximations and also fluctuations have been neglected. Hence it is not surprising that as $p$ increases,  the approximations affect the results significantly. Also, there are integrations which involve incomplete $\gamma$  function that can only be handled numerically which may add to the errors in the estimates of the terms. We have shown the analytical results for different $p$ values in Fig \ref{Analytical_1D_PN}.

\subsection{Distribution of relaxation time $T(x)$}
Since we are dealing with intermittent rest, it is important to study the relaxation time also. Whenever the forager hits a food (say at $x$) it may rest (relax) for some time $T$, we denote the relaxation time at $x$ as $T(x)$. The detailed analysis of $T(x)$ will be discussed in this section.

\begin{figure}[h!]
	\begin{center}
		\includegraphics[angle=-90, trim = 0 0 0 0, clip = true, width=0.99\linewidth]{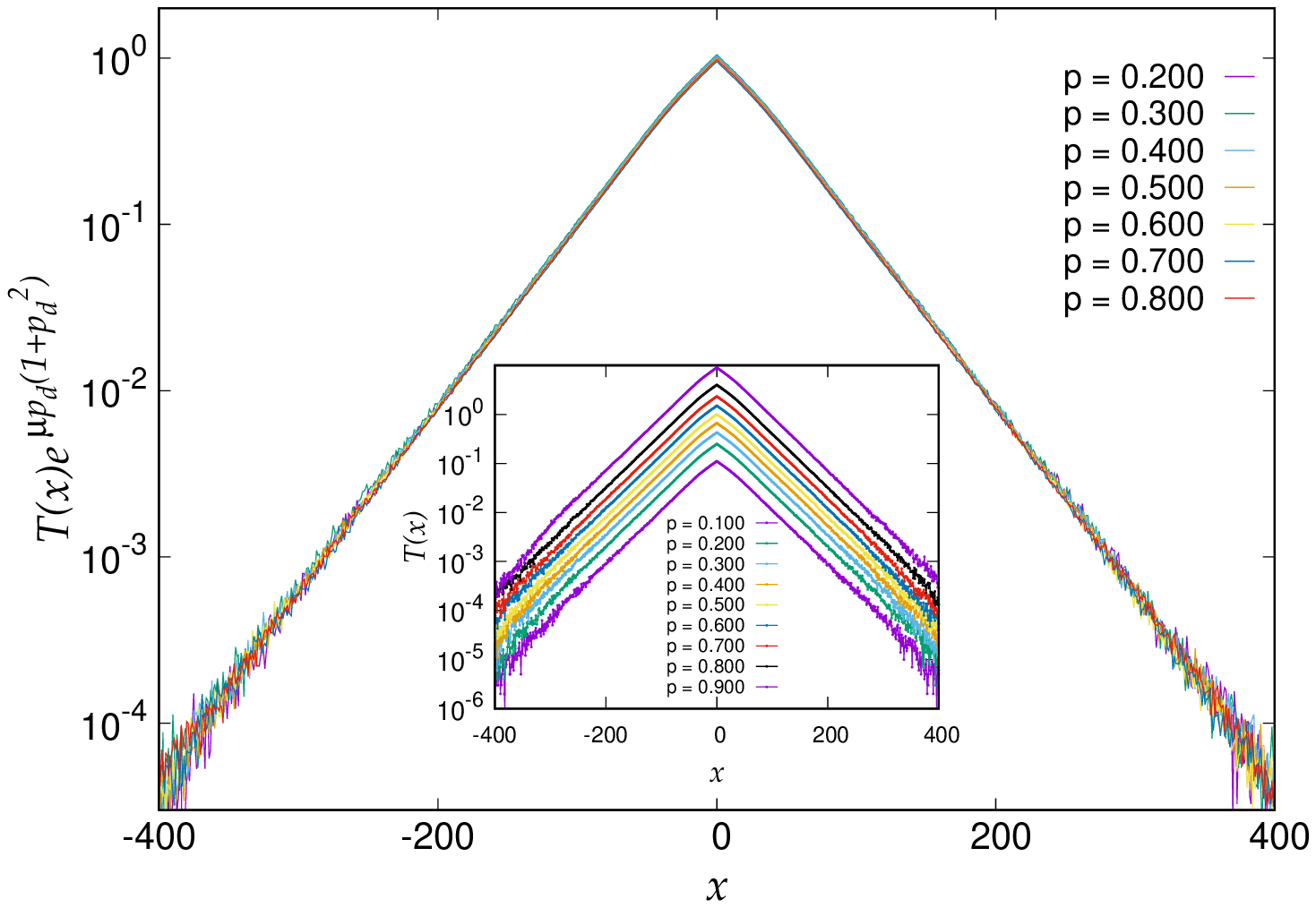}
		\caption{Behavior of relaxation time $T(x)$ over sites $x$ (in $1d$). Data collapse of $T(x)e^{\mu p_d(1+p_d^2)}$ against $x$ has been shown with $0.20 \leqslant p \leqslant 0.80$ with $\mu = 4.3$ and $p_d = 0.5 - p$. In the inset plot of relaxation time $T(x)$ against sites $x$ has shown.}
		\label{1D_T_x}
	\end{center}
\end{figure}

The plot of average relaxation time $T(x)$ versus $x$ shows a maximum at the origin. Due to the initial condition the forager always has a finite probability to take rest at origin irrespective of $p$. It has been observed in Fig. \ref{1D_T_x} (shown in the inset) that the relaxation time follows the relation, 
\begin{equation} \label{T}
	T(x) \propto \exp(-\kappa_T|x|).
\end{equation} 
The curves show symmetry and as $p$ increases, $T(x)$ also increases. The data collapse for the same has also been shown in Fig. \ref{1D_T_x}.
\\

\begin{figure}[h!]
	\begin{center}
		\includegraphics[angle=-90, trim = 0 0 0 0, clip = true, width=0.99\linewidth]{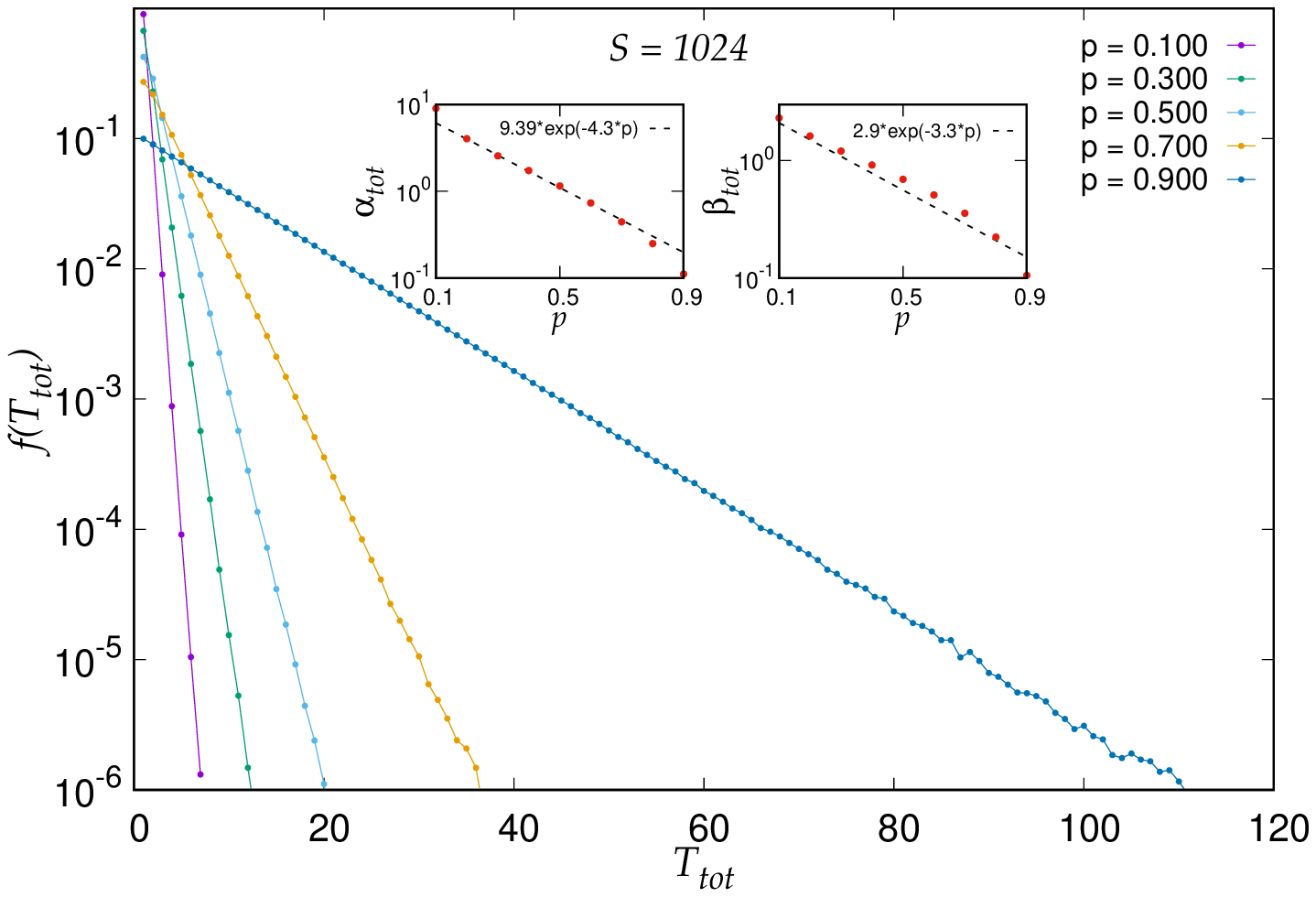}
		\caption{Distribution of relaxation time in $1d$ $f(T_{\text{tot}})$  against $T_{\text{tot}}$ with $0.1 \leqslant p \leqslant 0.9$ and $S = 1024$. In the inset, the variation of $\alpha_{\text{tot}}$ and $\beta_{\text{tot}}$ have been shown.}
		\label{PT_1D}
	\end{center}
\end{figure}

For a single realization, the total relaxation time is defined as $T_{\text{tot}}$, obtained by summing the relaxation times over all sites. After obtaining $T_{\text{tot}}$ for each realization, the configuration average is computed. We denote the distribution of $T_{\text{tot}}$ as $f(T_{\text{tot}})$. We have calculated $f(T_{\text{tot}})$ for $S = 1024$ in $1d$ as well as in $2d$. In $1d$, the behavior is shown in Fig. \ref{PT_1D}. 
It appears that the curve follows the relation, 

\begin{equation} \label{PT}
	f(T_{\text{tot}}) = \alpha_{\text{tot}}(p) \exp(-\beta_{\text{tot}}(p) T_{\text{tot}}),
\end{equation}
where $\alpha_{\text{tot}}$ and $\beta_{\text{tot}}$ also decay exponentially.
The variation of $\alpha_{\text{tot}}$ and $\beta_{\text{tot}}$ are shown in the inset of Fig. \ref{PT_1D}. It has been found in 1$d$ that $\alpha_{\text{tot}}(p) = 9.23\exp(-4.2p)$ and $\beta_{\text{tot}}(p) = 3.18\exp(-3.2p)$, whereas for 2$d$ case, $\alpha_{\text{tot}}(p) = 11.2\exp(-4.5p)$ and $\beta_{\text{tot}}(p) = 3.12\exp(-3.1p)$. 

\subsection{Distribution of inter-encounter time $P_t(N_t)$}
We define $N_t$ as the time interval between two successive food encounters by the forager. At each encounter with food, the forager's internal energy resets to full. Such  intervals, collected until starvation, yields the distribution $P_t(N_t)$. 

The statistical properties of $P_t(N_t)$ provide a detailed characterization of the foraging dynamics, reflecting the interplay among the forager's motion, environmental structure and food availability. 

\begin{figure}[h!]
	\begin{center}
		\includegraphics[angle=-90, trim = 0 0 0 0, clip = true, width=0.99\linewidth]{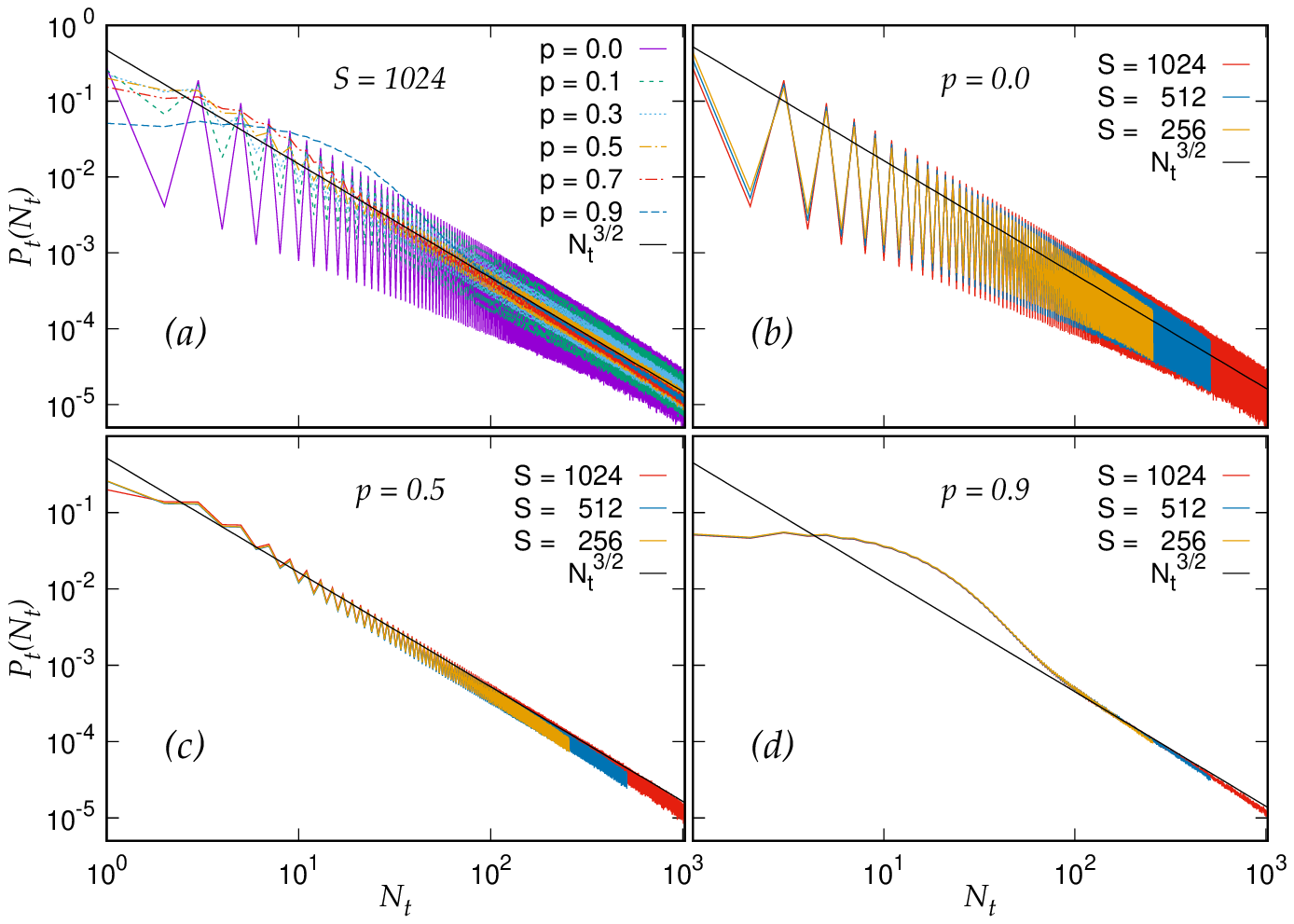}
		\caption{Plot of distribution $P_t(N_t)$ of inter-encounter time $N_t$ for the following : (a) varying $p$ for $S=1024$, (b) varying $S$ for $p=0$, (c) varying $S$ for $p=0.5$ and (d) varying $S$ for $p=0.9$.}
		\label{1D_Pt_collage}
	\end{center}
\end{figure}

It is clear that $P_t(N_t)$ depends on $N_t$ only,  for any $p$. For small $p$, the dependence is  $N_t^{-1.5}$ over almost the entire region and for larger $p$, only for large $N_t$.
This should be related to the time interval between visiting two sites, both for the first time, for a diffusing walker as follows:
It is well known that the First passage probability $F(t)  \propto 1/t^{3/2}$ for a diffusing walker at large time $t$ \cite{RednerFP,Basu}. 
Consider $t_1$ and $t_2$ be the times when visits are made for the first time and in between these two, all visits are not first time visits. We have $t_2 - t_1 = N_t$. 
Thus we have,
\begin{equation}
    P_t(N_t) = F(t_1) \bigg[\prod_{i=1}^{t-1} (1- F(t_1 + i))\bigg] F(t_1+N_t),
\end{equation}
which can be calculated easily as the product term approaches $1$ for large time and therefore the dependence is $N_t^{-3/2}$ from the last term. This approach is of course correct when 
most of the time the forager performs usual random walk without resting. Thus, the results match well for 
$p$ up to $0.5$ and also for larger values of $N_t$ when sites with food 
become rare and the forager is forced not to take rest most of the time. Beyond this $p$ value, the resting contributes more to the dynamics, but for smaller values of $N_t$ only, when more sites with 
food are available. As a result, a shallow peak occurs for larger values of $p$ at an intermediate value of $N_t$. 
The distributions are shown in Fig. \ref{1D_Pt_collage}. In addition, there is a zig-zag pattern in the distribution for small values of $p$, which can be explained as follows. For $p=0$, the forager moves without any resting; therefore, if it starts from either an even or an odd site in 1$d$, it can reach respectively an even or an odd site after even number of steps. Similarly, if the number of steps is odd, it lands on an odd (even) site, starting from an even (odd) site. When the forager finds food on a site,  it means that that site is visited for the first time. In one dimension, there will be one and only one site adjacent to this site which will be definitely non-empty or unvisited. Since this site can be reached in an odd number of steps, there will be a bias towards odd values of $N_t$, the interval of time between two successive food encounters. This leads to the zigzag pattern in $P_t(N_t)$. However, when $p>0$, intermittent resting destroys the even-odd structure of the walk and the distribution becomes smooth.



\section{Conclusions}

In this work, we studied a forager model on a $d$-dimensional lattice with
starvation constraints and introduced a novel element of intermittent rest such that  the
forager may rest with probability $p$ upon encountering food. The food is
distributed uniformly on the lattice, and the forager can
survive for $S$ steps without consuming any food further. We analyzed the system's  behavior by focusing on the key observables as follows: the average lifetime $\tau$, the average
number of distinct sites visited $N$ and the relaxation time $T(x)$ associated
with a site $x$.
Our results reveal rich dynamical features that strongly depend on both
the starvation time $S$ and the parameter for intermittent rest $p$. For $p=0$, our studies for $\tau$ and $N$ align quite well with existing studies. For $p=0$, the known results have been 
reproduced both in  one  and two dimensions. The case $p=1$ yields trivial behavior, corresponding to a stationary forager with minimal survival. However, for intermediate values $0 < p < 1$, the system exhibits non-linear crossovers and new scaling regimes. The behaviors of scaled lifetimes $\tau/S^{\Delta}$, with $\Delta \simeq 1, 2$ for $1d$ and $2d$ respectively, are also studied which shows a maxima for a given $0 < p < 1$ at some $S_{\text{max}}$. Specifically, we identify three characteristic scales for $\tau$, $N$ and  $\tau/S^{\Delta}$ respectively as $S_{\text{crossing}}$, $S_{\text{merge}}$ and $S_{\text{max}}$, the first two being proportional to $1/(1-p)$ and the third to $1/{(1-p)}^z$ ($z$ is approximately $1$), beyond which the forager’s behavior aligns with the $p=0$ regime. These indicate that there is a strong influence of intermittent resting on the exploration dynamics.
We further established a scaling relation between he lifetime and the number of distinct
sites. A change in the scaling behavior has been observed below and above a particular $N$ value, viz., $N^*$; below $N^*$, $\tau \sim N$ and above $N^*$, $\tau \sim N^2$ for 1$d$. The values $N^*$ for different $p$ are compared to the values of $N$ at $S_{\text{merge}}$ in Table \ref{Nstar} and they are quite similar, especially for low $p$. Thus, $\tau(N)=\alpha(p)N^{\beta}$, with $\beta=1$ or $2$ as mentioned above, and $\alpha(p)=1/(1-p)$. It is worth mentioning here that, all the results for $\tau$ and $N$ as in \cref{A,B,C} describe an extension of lifetime of the forager when $S$ and $p$ are high.

The relaxation time $T(x)$ was found to decay exponentially with $x$. The distribution of $T_{\text{tot}}$, summed over $x$, i.e., $P(T_{\text{tot}})$ follows an exponential form, with parameters dependent on $p$. Additionally, we studied the distribution of inter-encounter times, which further substantiates the model’s stochastic character.

All the results show that significant changes occur when $p > 0.5$ where the results clearly deviate from those at $p=0$, at least when $S$ is finite. This suggests that there is a threshold value of $p$ above which the effect of the intermittent rest becomes non trivial. However, it is difficult to exactly detect the threshold value.

Altogether, this study demonstrates that the inclusion of intermittent resting, an
elementary yet biologically motivated modification, introduces significant
changes to the statistical features of classical starving random walk models.
These findings may be useful in understanding real-world foraging behavior, intermittent search processes, and transport phenomena in disordered
systems.

In future, incorporating memory or adaptive strategies may enrich the model further, offering connections to a more realistic situation.

The code used to generate the data is available in the GitHub
repository \cite{AquibRepo}.

The authors thank Prof. Sidney Redner for some delightful comments and discussions. AM acknowledges financial support from CSIR, India (Grant no. 08/0463(12870)/2021-EMR-I). AM and SG acknowledge
the computational facility of Vidyasagar College, University of Calcutta. 
PS acknowledges financial support from CSIR scheme 03/1495/23/EMR-II. 

\onecolumngrid
\appendix

\section{Detailed Calculations for $P(N)$} 
To account for intermittent resting in the continuum description, we introduce the parameter $p$ and write the corresponding modified diffusion equation in the following form.
\begin{equation}
	\frac{\partial u(x,t)}{\partial t} = \left\{1-p\Bigl(1 - \Bigl \langle\frac{\eta_t}{\eta}\Bigl \rangle\Bigr)\right\}D \frac{\partial^2 u(x,t)}{\partial x^2}.
    \label{DiffusionEquation}
\end{equation}
Here, $\eta_t$ denotes the number of distinct sites visited up to time $t$, and $\eta$ is the total number of visited sites at starvation  and $1-\eta_t/\eta$ can be interpreted as the probability of visiting a so far unvisited site. Since the ratio $\eta_t/\eta$ fluctuates strongly across realizations, we use its ensemble-averaged form $\braket{\eta_t/\eta}$, which yields a smooth and representative quantity suitable for the continuum formulation.The form of this equation is chosen such that it reduces to the ordinary diffusion equation in the limits $p=0$ and $\Bigl \langle \frac{\eta_t}{\eta} \Bigl \rangle \rightarrow 1$. 

We use the functional form $\langle \eta_t / \eta \rangle = (t/(t+\Theta))^{\zeta(p)}$ from the behavior of $\braket{\eta_t/\eta}$. The exponent $\zeta(p)$ can be obtained from numerical fitting of the curve. The form of $\braket{\eta_t/\eta}$ is deliberately chosen so that for $t \gg \Theta$ the quantity saturates to unity. In particular, we found for $p=0$, $\zeta(0)=1/2$, corresponding to the efficient exploration regime, while in the limit $p \to 1$, $\zeta(p)\to1$. 
The above equation can then be written as,
\begin{equation}
	\frac{\partial u(x,t)}{\partial t} = \Biggl[1-p\biggl\{1 - \Bigl(\frac{t}{t+\Theta}\Bigr)^{\zeta(p)}\biggr\}\Bigg]D \frac{\partial^2 u(x,t)}{\partial x^2}. \label{DiffusionEquationA2}
\end{equation}



Using the method of separation of variable $u(x,t) = X(x)T(t)$, we get,\\

\begin{equation}
	\frac{1}{\biggl[1-p\Big\{1 - \left(\frac{t}{t+\Theta}\right)^{\zeta}\Big\}\biggr]D}\frac{1}{T} \frac{dT(t)}{dt} = \frac{1}{X}\frac{d^2X}{dx^2} = -\frac{\lambda^2}{a^2}.
\end{equation}
where, $\lambda =$ constant and $a = $ lattice spacing,
\begin{equation}
	u(x,t) =  \left[ A \sin\Big(\frac{\lambda x}{a}\Big) + B \cos\Big(\frac{\lambda x}{a}\Big) \right] \exp\Bigg[-\bigg\{\Big(\frac{\lambda}{a}\Big)^2D \bigg\}\int dt \bigg\{(1-p) + p\Big(\frac{t}{t+\Theta}\Big)^{\zeta}\bigg\}\Bigg].
\end{equation}
Assuming $t \gg \Theta$,
\begin{equation}
	u(x,t) = \left[ A \sin\Big(\frac{\lambda x}{a}\Big) + B \cos\Big(\frac{\lambda x}{a}\Big) \right] \Bigg[ t^{\left(\frac{\lambda }{a}\right)^2D\Theta p\zeta}\exp\left\{-\Big(\frac{\lambda }{a}\Big)^2Dt \right\} \Bigg].
\end{equation}
Boundary conditions are,\\
$$u(0,t) = u(ja,t) = 0$$ (since the walker stops it $ja$)\\
We get $B = 0$ and $\lambda_n = n\pi/j$,
\begin{equation}
	u(x,t) = \sum_{n=1}^{\infty} A_n \sin\Big(\frac{n\pi x}{ja}\Big) \Bigg[ t^{\left(\frac{n\pi }{ja}\right)^2D\Theta p\zeta}\exp\left\{-\Big(\frac{n\pi }{ja}\Big)^2Dt \right\} \Bigg].
\end{equation}
To find the value of $A_n$ we will use Fourier trick and the initial condition
$u(x,0) = \delta(x-a)$ (since it starts from a distance $a$)
\begin{eqnarray}
	u(x,0) & = & \delta(x-a) \nonumber \\ 
	\sum A_n \sin\Big(\frac{n\pi x}{ja}\Big) & = & \delta(x-a) \nonumber \\
	\sum \int_0^{ja} dx A_n \sin^2\Big(\frac{n\pi x}{ja}\Big) & = & \int_0^{ja} dx \text{  }  \delta(x-a) \sin\Big(\frac{n\pi x}{ja}\Big)\nonumber \\
	A_n & = & \frac{2}{ja}\sin\Big(\frac{n\pi}{j}\Big).
\end{eqnarray}
Therefore, 
\begin{equation}
	u(x,t) = \frac{2}{ja} \sum_{n=1}^{\infty} \sin\Big(\frac{n\pi}{j}\Big) \sin\Big(\frac{n\pi x}{ja}\Big) \Bigg[ t^{\left(\frac{n\pi }{ja}\right)^2D\Theta p\zeta}\exp\left\{-\Big(\frac{n\pi }{ja}\Big)^2Dt \right\} \Bigg].
\end{equation}

Instead of performing the integral, we calculate the flux across the boundary $[0,ja]$.

\begin{align}
	F_j(t) & =  D\left( \frac{\partial u}{\partial x}\bigg|_{x = 0} - \frac{\partial u}{\partial x}\bigg|_{x = ja}\right) \nonumber \\
	& = \frac{2D}{ja} \sum_{n=1}^{\infty} \frac{n\pi}{ja}  \sin\left(\frac{n\pi}{j} \right) \left( 1 - (-1)^n \right)\left[ t^{\left(\frac{n\pi }{ja}\right)^2D\Theta p\zeta}\exp\left\{-\left(\frac{n\pi }{ja}\right)^2Dt \right\} \right] \nonumber\\
	F_j(t)& = \frac{4\pi D}{(ja)^2} \sum_{n = 1}^{\infty} (2n+1)\sin{\left( \frac{(2n+1)\pi}{j}\right)}\left[ t^{\left(\frac{(2n+1)\pi }{ja}\right)^2D\Theta p\zeta}\exp\left\{-\left(\frac{(2n+1)\pi }{ja}\right)^2Dt \right\} \right]. \label{eqn: F_j}
\end{align}

Now, $Q_j = \int_0^{S} dt F_j(t)$. \\
Assuming total flux that can escape is normalized to 1, we may write, $\lim_{S \rightarrow \infty} Q_j = 1$.\\

Therefore, 

\begin{align}
	Q_j & = 1 - \int_S^{\infty} dt F_j(t) \nonumber \\
	& = 1 - \frac{4\pi D}{(ja)^2} \sum_{n = 1}^{\infty} (2n+1)\sin{\left( \frac{(2n+1)\pi}{j}\right)}\int_S^{\infty} dt \left[ t^{\left(\frac{(2n+1)\pi }{ja}\right)^2D\Theta p\zeta}\exp\left\{-\left(\frac{(2n+1)\pi }{ja}\right)^2Dt \right\} \right] \nonumber \\
	& = 1 - \frac{4\pi D}{(ja)^2} \sum_{n = 1}^{\infty} (2n+1)\sin{\left( \frac{(2n+1)\pi}{j}\right)} \frac{\gamma\left[ \left(\frac{(2n+1)\pi }{ja}\right)^2D\Theta p\zeta + 1 , \left(\frac{(2n+1)\pi }{ja}\right)^2DS\right]}{\left[\left(\frac{(2n+1)\pi }{ja}\right)^2D\right]^{\left(\frac{(2n+1)\pi }{ja}\right)^2D\Theta p\zeta + 1}} \nonumber \\
	Q_j & = 1 - \frac{4}{\pi} \sum_{n = 1}^{\infty} \frac{\sin{\left( \frac{(2n+1)\pi}{j}\right)}}{(2n+1)} \frac{\gamma\left[ \left(\frac{(2n+1)\pi }{ja}\right)^2D\Theta p\zeta + 1 , \left(\frac{(2n+1)\pi }{ja}\right)^2DS\right]}{\left[\left(\frac{(2n+1)\pi }{ja}\right)^2D\right]^{\left(\frac{(2n+1)\pi }{ja}\right)^2D\Theta p\zeta}}. \label{generalQ_j}
\end{align}
where, 
\begin{equation}
    \gamma (s,x)= \int_x^{\infty} dt \text{ } t^{s-1}e^{-t} \label{gamma_fn}
\end{equation} 
The probability distribution of average number of distinct visited site $N$, i.e., $P(N)$,
\begin{equation}
	P(N) = Q_2Q_3 \cdots Q_{N}(1-Q_{N+1}), \label{PN}
\end{equation}
where $(1-Q_{N+1})$ signifies the exemption of $(N+1)$th site.\\
To calculate the explicit result we need to calculate the product of $Q_j$'s,
\begin{align}
	U_{N} & = \prod_{2\leqslant j \leqslant N} Q_j \nonumber\\
    \ln{U_N} & = \sum_{j = 2}^{N} \ln{\left[ 1 - \frac{4}{\pi} \sum_{n = 1}^{\infty} \frac{\sin{\left( \frac{(2n+1)\pi}{j}\right)}}{(2n+1)} \frac{\gamma\left[ \left(\frac{(2n+1)\pi }{ja}\right)^2D\Theta p\zeta + 1 , \left(\frac{(2n+1)\pi }{ja}\right)^2DS\right]}{\left[\left(\frac{(2n+1)\pi }{ja}\right)^2D\right]^{\left(\frac{(2n+1)\pi }{ja}\right)^2D\Theta p\zeta}}  \right]}. \label{logUN}
\end{align}

Again from Eq. \ref{generalQ_j}, for $j = N+1$, 

\begin{equation}
    1 - Q_{N+1}  = \frac{4}{\pi} \sum_{n = 1}^{\infty} \frac{\sin{\left( \frac{(2n+1)\pi}{N+1}\right)}}{(2n+1)} \frac{\gamma\left[ \left(\frac{(2n+1)\pi }{(N+1)a}\right)^2D\Theta p\zeta + 1 , \left(\frac{(2n+1)\pi }{(N+1)a}\right)^2DS\right]}{\left[\left(\frac{(2n+1)\pi }{(N+1)a}\right)^2D\right]^{\left(\frac{(2n+1)\pi }{(N+1)a}\right)^2D\Theta p\zeta}} .\label{QN+1}
\end{equation}
 
Now, substituting Eq. \ref{logUN} and \ref{QN+1} in Eq. \ref{PN} we get a general form of $P(N)$.

\begin{multline}
	P(N) = \frac{4}{\pi} \sum_{m = 1}^{\infty} \frac{\sin{\left( \frac{(2m+1)\pi}{N+1}\right)}}{(2m+1)} \frac{\gamma\left[ \left(\frac{(2m+1)\pi }{(N+1)a}\right)^2D\Theta p\zeta + 1 , \left(\frac{(2m+1)\pi }{(N+1)a}\right)^2DS\right]}{\left[\left(\frac{(2m+1)\pi }{(N+1)a}\right)^2D\right]^{\left(\frac{(2m+1)\pi }{(N+1)a}\right)^2D\Theta p\zeta}} \\
	\times \exp{\left[ \sum_{2 \leqslant j \leqslant N} \ln{\left[ 1 - \frac{4}{\pi} \sum_{n = 1}^{\infty} \frac{\sin{\left( \frac{(2n+1)\pi}{j}\right)}}{(2n+1)} \frac{\gamma\left[ \left(\frac{(2n+1)\pi }{ja}\right)^2D\Theta p\zeta + 1 , \left(\frac{(2n+1)\pi }{ja}\right)^2DS\right]}{\left[\left(\frac{(2n+1)\pi }{ja}\right)^2D\right]^{\left(\frac{(2n+1)\pi }{ja}\right)^2D\Theta p\zeta}}  \right]}\right]}. \label{generalPN_repeat}
\end{multline}
Obtaining a closed form expression of $P(N)$ from the above equation is difficult as one has to  deal with the incomplete $\gamma$ function in Eq. \ref{generalPN_repeat}. To calculate it numerically,  the value of $\Theta$ has to be determined. However, if $\Theta$ is large, then even a numerical 
estimation becomes impossible. 
We found that for smaller value of $p$, Eq. \ref{generalPN_repeat}, using $\Theta=1$ gives results that match the simulation results quite well as shown  in Fig. \ref{1D_PN}. These results are not much dependent on $\Theta$ as $\Theta$, associated with the $p$ dependent part of \ref{DiffusionEquation} does not affect the low $p$ regime. For high $p$ values, however, the results are much more sensitive to 
the value of $\Theta$ and there is a significant deviation of the analytical results when compared to numerical simulation results (shown for $\Theta =1$ in Fig. \ref{Analytical_1D_PN}).


Setting $p= 0$ in Eq. \ref{generalPN_repeat} we get,
\begin{equation}
    P(N) \simeq \frac{4}{N} \sum_{m = 1}^{\infty} \exp{\left[-\frac{(2m+1)^2 \pi^2DS}{N^2a^2} -2 \sum_{n = 1}^{\infty} E_1\left( \frac{(2n+1)^2\pi^2DS}{N^2a^2} \right)\right]}. \label{analitical_PN}
\end{equation}\\
where, $E_1(x) = \int_1^{\infty} dt \frac{e^{-xt}}{t}$.\\ \\ Equation \ref{analitical_PN} exactly matches with the result of Ref.\cite{Redner2014}.

\twocolumngrid

\end{document}